# The Cosmic Hitchhikers Hypothesis: Extraterrestrial Civilizations Using Free-Floating Planets for Interstellar Colonization

Irina K. Romanovskaya

E-mail: irina.mullins@hccs.edu

January 5, 2022

**Abstract**

I propose the Cosmic Hitchhikers hypothesis as follows. Advanced extraterrestrial civilizations may use free-floating planets as interstellar transportation for space exploration and interstellar colonization. Large groups or populations of their biological species, post-biological species, and technologies may become Cosmic Hitchhikers when they ride free-floating planets to reach, explore and colonize planetary systems. To get an interstellar ride, Cosmic Hitchhikers may travel to free-floating planets passing close by their home worlds. Otherwise, they may use astronomical engineering to steer free-floating planets toward their home planetary systems. Cosmic Hitchhikers may also ride objects native to the outer regions of their planetary systems, which become free-floating planets when ejected by astronomical engineering or by their stars during the asymptotic giant branch evolution. During interstellar travel, Cosmic Hitchhikers may apply astronomical engineering to steer their free-floating planets toward the planetary systems of their choice. Whereas riding free-floating planets may not save travel time, it avoids the technical challenges of interstellar spacecraft transporting large populations. Each civilization of Cosmic Hitchhikers may colonize several planetary systems. Its colonies may grow into autonomous civilizations, changing the number of civilizations in the Galaxy. Over the last 4 billion years, Cosmic Hitchhikers or their artifacts riding free-floating planets might have passed by the Solar System. Therefore, their artifacts might exist in the Solar System or in our stellar neighborhood. SETI and SETA should include the search for Cosmic Hitchhikers and their artifacts. Keywords: SETI, SETA, free-floating planet, extraterrestrial civilization, interstellar travel, interstellar colonization, artifact, Cosmic Hitchhikers

# 1. Introduction

There are many reasons for why interstellar spacecraft travelling can become an unsuccessful endeavor for large populations of biological and post-biological species. For example, space travelers may run out of consumable resources and means of maintaining and repairing their spacecraft before they arrive in planetary systems. A relativistic spacecraft may be negatively affected by its interactions with gas and dust in the interstellar medium (Hoang *et al*., 2017).

I propose that free-floating planets, which are planetary-mass objects that are not gravitationally bound to stars, can be used as a means of interstellar travel for large groups and populations of intelligent biological and post-biological species as well as their technologies. Also known as rogue or nomad planets, free-floating planets may have a liquid ocean under a thick atmosphere or an ice layer, and some free-floating planets may host simple life forms, especially in subsurface environments (Abbot and Switzer 2011; Stevenson 1999; Badescu 2011; Lingam and Loeb 2019). Several models theoretically predicted that some exomoons orbiting free-floating planets may retain an atmosphere and liquid water on their surface (Ávila *et al*., 2021). Furthermore, studies suggest that the number of free-floating planets may be substantial in our Galaxy (Sumi *et al*. 2011; Caballero 2018; Safonova and Sivaram 2019; Strigari *et al*. 2012; Dai and Guerras 2018).

I propose the Cosmic Hitchhikers hypothesis according to which large groups and populations of biological species, post-biological species, and technologies of advanced extraterrestrial civilizations may become Cosmic Hitchhikers when they travel from planetary systems to free-floating planets and use such free-floating planets as interstellar transportation to reach, explore and colonize other planetary systems. I discuss advantages of using free-floating planets as a means of interstellar travel for space exploration and interstellar colonization. I recommend that SETI and SETA should include the search for Cosmic Hitchhikers and their artifacts.

# 2. Free-Floating Planets: Are They Abundant or Rare?

Thousands of exoplanets gravitationally bound to their host stars have been discovered in our Galaxy so far. However, astronomical observations have discovered a limited number of free-floating planets because their detection remains to be very challenging. Gravitational microlensing, a method used to search for gravitationally bound and unbound exoplanets, requires a special and rare condition of alignment between a free-floating planet and a background star during observations. Infrared imaging surveys have been used to discover free-floating planets with high atmospheric temperatures. For example, infrared imaging surveys discovered free-floating planets with mass probably as small as a few times the mass of Jupiter and with atmospheric effective temperatures of 1700 to 2200 K residing in young star-forming region (Osoriov et al., 2000). Miret-Roig and her colleagues reported a discovery of a rich population of Jupiter-like free-floating planets (between 70 and 170 free-floating planets) in the Upper Scorpius young stellar association. These young planets are still hot enough to glow, so that they can be detected in optical and near-infrared wavelengths (Miret-Roig N *et al.*, 2021).

Researchers have different opinions on the number of free-floating planets in the Galaxy, as detection of free-floating planets remains difficult, and current theories describing formation of free-floating planets vary because of a lack of large homogeneous samples needed for a statistical analysis of their properties. Free-floating planets may originally form around a host star and in multiple-star systems, and then they may be scattered away. Alternatively, a free-floating planet may form in isolation through direct collapse of a cloud of gas and dust, similarly to star formation (Gahm *et al.*, 2007). Potentially, both isolated formation of free-floating planets from clouds of gas and dust as well as their formation when planets are scattered from planetary systems may contribute to the population of free-floating planets.

If free-floating planets form in common cosmic events, then their number in the Galaxy should be very large. This is similar to the reasoning that the existence of a great number of exoplanets orbiting stars in our Galaxy, is linked to the proven existence of disks that orbit young stars, as well as theories and observations of the formation of planets in such disks.

Currently, there are estimates indicating that our Galaxy contains a significant number of free-floating planets. For example, Barclay and his colleagues ran 300 N-body simulations of terrestrial planet formation around a solar-type star, with and without giant planets present. Their study showed that about 2.5 terrestrial-mass planets per star become free-floating planets after they are ejected during the planet formation process. The population of such free-floating planets is likely composed of Mars-sized planets (Barclay et al., 2017). Other studies predicted that there may be many Jupiter-mass free-floating planets. For example, Sumi and his colleagues used two years of gravitational microlensing survey observations toward the Galactic Bulge to estimate that there are two Jupiter-mass free-floating planets per each main-sequence star (Sumi et al. 2011), though their estimate may be re-evaluated. One study predicts that per each main-sequence star in our Galaxy, there may be up to $10^5$ compact objects in the mass range $10^{-8}$–$10^{-2}$ solar mass that are not gravitationally bound to stars (Strigari *et al.*, 2012).

Free-floating planets, which initially form in a disk orbiting a star, can become unbound thanks to various common processes. For example, free-floating planets may be produced in the process of ejection of fragments from a protoplanetary disk when it is perturbed (Vorobyov and Pavlyuchenkov, 2017). Planets can be also ejected by interactions with another star (Hurley & Shara 2002). Post-main-sequence stars can eject some planets that orbit them: (a) Oort Clouds and wide-separation planets may be dynamically ejected from 1–7 times Solar mass parent stars during the asymptotic giant branch (AGB) evolution; (b) most of the planetary material that survives a supernova from a 7–20 times Solar mass progenitor will become dynamically ejected from the system; (c) planets orbiting >20 times Solar mass black hole progenitors may survive or be ejected (Veras *et al.*, 2011).

Free-floating planets can be ejected by scattering interactions in a multi-planet system (Veras & Raymond 2012). Planets are also more susceptible to ejection from multiple-star planetary systems than from single-star planetary systems for a given system mass (Veras and Tout, 2012). According to Veras and Tout, multiple stars in multiple-star planetary systems can violently interact in ways that a single evolving star cannot. Therefore, the effect on orbiting them objects is greater than that in the single-star case.

Veras and Tout conservatively estimated that: (a) planetary material, which is located beyond a few hundred AU while orbiting multiple stars each more massive than the Sun and whose minimum separation is less than 100 solar radii, is likely to be ejected during post-main-sequence evolution of the stars; (b) all Oort cloud analogues in post-main-sequence multiple-star systems would be disrupted and could escape; (c) planets residing at a few tens of AU from the central concentration of stars may escape. Veras and Tout proposed that these systems may significantly contribute to the free-floating planet population (Veras and Tout, 2012). It also follows from the studies conducted by Smullen and her colleagues that if planet formation around binary stars is very efficient, then circumbinary planetary systems might be responsible for producing free-floating planets (Smullen et al., 2016).

Additionally, studies proposed that the disruption of a binary star system by the massive black hole at the Galactic Centre, SgrA*, could result in the capture of one star around SgrA* and the ejection of its companion star as a hypervelocity star. If the binary system would have a planet, then for some orbital parameters, the planet could be ejected at a high speed and it would travel as a hypervelocity free-floating planet (Ginsburg et al., 2012).

## 3. The Cosmic Hitchhikers Hypothesis

The Cosmic Hitchhikers hypothesis relies on two assumptions: (1) that the Copernican principle is valid, and, therefore, our Galaxy hosts more than one spacefaring civilization, and (2) that a significant number of free-floating planets exist in the disk of the Galaxy. The second assumption relies on observations, computer simulations and theories of free-floating planets, which I described earlier in this paper.

I propose the Cosmic Hitchhikers hypothesis as follows. Spacefaring extraterrestrial civilizations may use free-floating planets as a means of interstellar travel for space exploration and interstellar colonization. Large groups and populations of their biological species, post-biological species, and technologies become Cosmic Hitchhikers when they ride free-floating planets to reach, explore and colonize planetary systems. To get an interstellar ride, Cosmic Hitchhikers may travel to free-floating planets when the free-floating planets pass by their home worlds. Otherwise, they may use astronomical engineering to steer free-floating planets toward their home planetary systems. Cosmic Hitchhikers may also ride cosmic objects native to the outer regions of their planetary systems, which become free-floating planets after they are ejected by means of astronomical engineering or by their host stars during the asymptotic giant branch evolution. During interstellar travel, Cosmic Hitchhikers may use astronomical engineering to steer their free-floating planets toward the planetary systems of their choice. Cosmic Hitchhikers of one civilization may establish colonies in several planetary systems. The colonies may grow into autonomous civilizations, changing the total number of civilizations in the Galaxy. If Cosmic Hitchhikers exist, then over the last 4 billion years, Cosmic Hitchhikers or their artifacts riding free-floating planets might have passed by the Solar System. Therefore, their artifacts might exist in the Solar System or elsewhere in our stellar neighborhood.

The reason for spacefaring civilizations to use free-floating planets for interstellar travel would not be that of saving travel time. Space travel using free-floating planets would not take less time than space travel using spacecraft. Rather, the reason for using free-floating planets as a means of interstellar travel would be as follows. Using free-floating planets as interstellar transportation would allow extraterrestrial civilizations to avoid the technical (potentially unsolvable) challenges of interstellar spacecraft transporting large populations of species. Interstellar spacecraft travel could be most likely impossible for large groups or populations of spacefaring species because of the constraints placed on the number of species on board, the amounts of consumables their spacecraft could carry, the extent of protection from space radiation the spacecraft could provide, and the ability of the spacecraft to withstand interactions with interstellar travel environments negatively affecting its operations.

On the other hand, free-floating planets can provide large amounts of space and resources. Among other things, some free-floating planets with surface and subsurface oceans can provide water to be used as a consumable resource and for protection from space radiation. Travelers on free-floating planets would not have to worry about collisions with interstellar dust the way travelers on a relativistic spacecraft would have to worry.

Even if advanced extraterrestrial civilizations could build interstellar spacecraft for small to medium groups of their species, they could use free-floating planets to transport large groups or populations. For example, extraterrestrials could use free-floating planets to transport large groups or populations escaping oncoming existential threats, to misplace unwanted populations, to send large numbers of post-biological species to explore distant worlds, or to spread populations of their species to other planetary systems to preserve the continuity of their civilization (similar to how some people think of preserving the continuity of humankind by colonizing other planets). Extraterrestrial civilizations could also send Cosmic Hitchhikers in the form of their smart machines, probes, and other technologies to settle on free-floating planets and to conduct surveys of stars, planetary systems, and interstellar medium along the paths of the free-floating planets.

There may be at least four scenarios describing how Cosmic Hitchhikers might travel from their home worlds to free-floating planets. The first two scenarios involve free-floating planets passing by their home planetary systems, and the other two scenarios involve planets or planet-like objects being ejected from their home planetary systems.

*Scenario A: Using free-floating planets that pass by Cosmic Hitchhikers' home worlds*

Cosmic Hitchhikers may travel to free-floating planets when such planets pass by their home worlds. The probability of occurrence of such events would depend on the number and distribution of free-floating planets in the Galaxy, as well as the distribution of extraterrestrial civilizations. The distribution of free-floating planets and their dynamics depend on how free-floating planets originate. For example, it is considered that numerous free-floating planets should reside in stellar clusters. However, van Elteren and his colleagues ran computer simulations of the Orion Trapezium star cluster and concluded that about 80 percent of the free-floating planets would promptly escape the cluster upon being unbound from their host stars (van Elteren *et al.*, 2019).

Ginsburg and his colleagues proposed that the disruption of binary star systems by the massive black hole at the Galactic Centre, SgrA* can result in ejection of planets from such systems, and the ejected planets can travel from their binary stars and through the Galaxy as hypervelocity free-floating planets. (Ginsburg et al., 2012). Veras and his colleagues investigated relations between post-main-sequence stars and the fate of planets orbiting them. According to their studies, stars with 1–7 times solar mass undergoing the asymptotic giant branch evolution, as well as a supernova from a 7–20 times solar mass progenitor, can eject planets from their system (Veras *et al.*, 2011).

Therefore, free-floating planets can form in any region of the Galaxy where stars go supernova (if the planets survive their supernovae), where stars undergo the asymptotic giant branch (AGB) evolution, and where stellar clusters, as well as binary and multiple-star systems, may eject planets.

*Scenario B: Using free-floating planets steered toward Cosmic Hitchhikers' home worlds by means of astronomical engineering*

Cosmic Hitchhikers may use astronomical engineering to steer free-floating planets toward their home planetary systems. With regards to this possibility, human civilization can be used as an example. NASA sent astronauts to the Moon. For some time, NASA was considering sending astronauts to ride asteroids. NASA is also researching possible development of technologies that could change the direction of motion of asteroids in the Solar System.

When space agencies and privately owned enterprises investigate possibilities of using technologies to modify the motion of asteroids, they assert their intention to engage in astronomical engineering, which involves operations with the whole cosmic objects. For example, astronomical engineering methods of changing orbits of asteroids in the Solar System and sending them to the Moon or Mars for the purpose of mining were discussed by Misiak. Misiak proposed three possible scenarios of making an asteroid move in a certain direction: (a) sending a spacecraft equipped with nuclear weapon to the asteroid and setting nuclear explosion in front of the asteroid; (b) shooting the asteroid by 5-tons weapons from Earth's orbit, (3) attaching a huge solar sail to the asteroid (Misiak M, 2013).

Korycansky and his colleagues proposed to apply astronomical engineering strategy that uses gravitational assists to transfer orbital energy from Jupiter to Earth in order to modify the orbit of Earth and make Earth migrate farther away from the Sun (Korycansky D G, 2001). McInnes proposed astronomical engineering strategy that involves using a large reflective sail in order to produce a propulsive thrust caused by solar radiation pressure. If the sail were set to be in static equilibrium relative to Earth, then the center-of-mass of the Earth-sail system would slowly accelerate (McInnes C R, 2002). Badescu and Cathcart investigated hypothetical stellar engines that might be used to control to a certain extent the orbital motion of the Sun in the Galaxy (Badescu and Cathcart, 2006).

It is then reasonable to propose that advanced extraterrestrial civilizations could use astronomical engineering strategies to modify the motion of free-floating planets. Their strategies could involve sails driven by the pressure of electromagnetic radiation of some type as well as other methods and technologies. Astronomical engineering methods for modifying the motion of free-floating planets are unattainable to modern humans, but they may be realized by more advanced civilizations.

For example, a hypothetical extraterrestrial civilization, several centuries or a few thousand years more advanced than humankind, may be able to trace free-floating planets in its stellar neighborhood and to modify their motion similarly to how human space agencies learn to trace asteroids and how human scientists and engineers might eventually find the ways to control the motion of asteroids in the Solar System.

The advanced extraterrestrial civilization could have its automatic spacecraft exploring its stellar neighborhood, detecting and tracing free-floating planets. Its automatic spacecraft could detect a free-floating planet in the stellar neighborhood and send technologies to the free-floating planet that would steer it closer to the extraterrestrial civilization's planetary system so that its species could travel to it. It could be a long wait time for the extraterrestrials, but they could have reasons to do so. For example, they would want to escape an oncoming existential threat or to send large numbers of technologies or post-biological species that would explore distant worlds for millions of years.

*Scenario C: Using free-floating planets ejected from Cosmic Hitchhikers' home worlds by means of astronomical engineering*

Cosmic Hitchhikers may settle on cosmic objects native to the outer regions of their planetary systems and then use astronomical engineering to eject such objects from their planetary systems, thus artificially turning them into free-floating planets.

*Scenario D Using cosmic objects ejected from Cosmic Hitchhikers' home worlds by their host stars during the asymptotic giant branch evolution*

Cosmic Hitchhikers may ride cosmic objects native to the outer regions of their planetary systems, which become free-floating planets after they are ejected by their host stars during the asymptotic giant branch evolution.

For all the above scenarios, after reaching and settling on free-floating planets, Cosmic Hitchhikers could use astronomical engineering to modify the speed and direction of motion of their free-floating planets over the course of hundreds or thousands of years, so that the free-floating planets would bring the Cosmic Hitchhikers close to the planetary systems of their choice. During the long travel time, biological Cosmic Hitchhikers could have many generations developing new technologies and infrastructures. If Cosmic Hitchhikers were post-biological species and machines, they could simply deactivate themselves for a significant part of the duration of their interstellar travel.

Free-floating planets are a prime example of cosmic objects that Cosmic Hitchhikers may use as a means of interstellar travel. Which is why I focus on discussing how Cosmic Hitchhikers may use free-floating planets. Cosmic Hitchhikers could also use very large interstellar asteroids more similar to dwarf planets, if such interstellar asteroids exist. I leave out the discussion of whether such objects should be classified as free-floating planets or free-floating dwarf planets.

A civilization of Cosmic Hitchhikers using a free-floating planet as a means of interstellar travel could establish its colonies in more than one planetary system. Over time, the colonies could grow into independent civilizations, thus changing the total number of advanced civilizations in the Galaxy. Cosmic Hitchhikers in the form of automated probes could keep transferring probes from one free-floating planet to another, thus populating a growing number of free-floating planets and exploring the Galaxy for a long time.

If Cosmic Hitchhikers exist, then over the last 4 billion years, at least one free-floating planet carrying Cosmic Hitchhikers or their artifacts may have passed close by the Solar System or nearby planetary systems. The Cosmic Hitchhikers populating that free-floating planet may have potentially sent their technologies to the Solar System or to other nearby planetary systems. Therefore, artifacts of Cosmic Hitchhikers might exist in the Solar System or in our stellar neighborhood.

## 4. Using Spacecraft for Interstellar Travel versus Using Free-Floating Planets for Interstellar Travel

The ability to travel on a spacecraft to other stars is determined by the laws of mechanics, propulsion system, vehicle mass, and the means of life support and protection from space radiation. Interstellar spacecraft travel would require substantial amounts of consumables. Limitations on the amounts of consumables that spacecraft can carry, as well as inability to replenish consumable resources during interstellar travel, would limit the number of passengers. Interstellar travel environments would negatively affect spacecraft operations. These would include propulsive forces, space radiation, interstellar gas and dust, temperatures variations and more. As a result, spacecraft could become damaged or destroyed before arriving in other planetary systems. Hence, interstellar travel involving crewed spacecraft could be a futile endeavor for large groups or populations of space travelers, as they could run out of consumable resources and means of repairing their spacecraft before arriving in planetary systems.

Free-floating planets and their moons can be better suited as a means of interstellar travel for populations of intelligent biological and post-biological species as well as their technologies. Advantages of using free-floating planets and their moons as a means of interstellar travel for the purpose of interstellar colonization are discussed below.

*Advantage 1: Plentiful amounts of space for habitation and resources for in-situ resource utilization (ISRU)*

Free-floating planets or their moons may supply large amounts of resources and space for habitation, development and utilization of technologies. Free-floating planets may have a liquid ocean under a thick atmosphere or an ice layer sustained by radiogenic and primordial heat (Lingam and Loeb, 2020a). For example, free-floating planets with their composition and age similar to those of Earth and their mass of about 0.3 times Earth mass could maintain a liquid ocean under layers of water ice as a result of geothermal heat flux (Abbot and Switzer, 2011). Therefore, some free-floating planets may have environments capable of supporting simple life forms (Stevenson 1999; Badescu 2011; Abbot and Switzer 2011; Lingam and Loeb 2019). Theoretical studies predicted that some exomoons of free-floating planets may retain an atmosphere capable of creating conditions to ensure the long-term thermal stability of liquid water on their surface (Ávila *et al*., 2021).

*Advantage 2: Availability of liquid water that can be used for space radiation shielding*

Water can be used for space radiation shielding (DeWitt and Benton, 2020). If Cosmic Hitchhikers settle on free-floating planets or moons of free-floating planets with oceans of liquid water, then they may use that water for space radiation shielding. After developing and using technologies enabling colonization of oceans on free-floating planets and their moons, Cosmic Hitchhikers would also become better prepared for colonization of oceans in planetary systems.

*Advantage 3: Constant surface gravity*

Free-floating planets and moons of free-floating planets can provide constant surface gravity for interstellar travelers, even though it may differ from that of the travelers' home world. For multi-generational travel, extraterrestrial civilizations may apply biotechnologies to adapt to the surface gravity of free-floating planets or their moons.

*Advantage 4: Possibilities of applications of astronomical engineering*

Advanced civilizations may apply astronomical engineering to modify the motion of free-floating planets that they use as interstellar transportation.

As for interstellar spacecraft travel, Hansen and Zuckerman discussed one special case that involves extraterrestrial civilizations using spacecraft to travel to other stars that pass very close by the civilizations' home planetary systems. (Hansen and Zuckerman, 2021). Hansen and Zuckerman estimated that in the solar vicinity, one would expect appropriate close passages of other stars (about 100 times smaller than typical stellar separations) to occur at least once during a characteristic time of a Gyr.

I propose that free-floating planets as a means of relocation to other stars' planetary systems may have their advantages over spacecraft interstellar travel delivering civilizations to the closest stars. Namely, there is no guarantee that a star getting unusually close to an extraterrestrial civilization's home planetary system can offer a new safe home. This may be due to activity of the star or lack of suitable planets and moons in the habitable zone. The other star may also have its planetary system with its own life forms hostile to the civilization of newcomers.

On the other hand, free-floating planets are unbound and usually cold worlds that offer relatively stable environments. They may provide many generations of travelers with space, resources for In-Situ Research Utilization (ISRU) and protection from space radiation. Their motion in space may be altered by means of astronomical engineering. Populations of advanced civilizations may adapt their ways of living while riding free-floating planets, and they may have opportunities to decide which planetary systems to colonize.

Overall, different civilizations may choose different ways to send their populations to other stars, depending on their circumstances and technologies. If an advanced civilization would discover a G star approaching its home world closely, then the civilization could travel to it, and that star would become the civilization's new host star. If there are no suitable stars approaching the civilization within a reasonable waiting period, and the civilization has its means to get a ride on a free-floating planet toward planetary systems of its choice, then that civilization's populations might choose to become Cosmic Hitchhikers riding the free-floating planet.

## 5. Colonization of Free-Floating Planets versus Colonization of Planetary Systems

Among other reasons, spacefaring extraterrestrial civilizations could relocate their populations from their planetary systems to nearby free-floating planets and their moons when facing existential threats, such as artificially created disasters, cosmic violent events. However, free-floating planets may not serve as a permanent means of escape from existential threats, because even astronomical engineering modifying the motion of free-floating planets might not help extraterrestrials completely avoid all cosmic threats in the Galaxy.

Furthermore, because of the waning heat production in the interior of free-floating planets, such free-floating planets would eventually fail to sustain their oceans of liquid water (if the planets initially had them). Additionally, free-floating planets provide less resources than planetary systems.

Therefore, I hypothesize that instead of making free-floating planets their permanent homes, extraterrestrial civilizations would use free-floating planets as a means of interstellar travel to reach and colonize other planetary systems. In some cases, they could potentially reach and colonize planet-like objects orbiting brown dwarfs.

## 6. Space Colonization and the Number of Civilizations in the Galaxy

It is customary to hypothesize that an advanced civilization becomes a multiplanetary civilization after it colonizes planets (and moons). However, I argue that it is more likely that advanced civilizations do not become multiplanetary civilizations after they colonize planets and moons in one planetary system, or after they colonize more than one planetary system in the Galaxy. Instead, when species of an advanced civilization (i.e., a *parent-civilization*) establish colonies on other cosmic objects of their home planetary system and in other planetary systems, their colonies become the "seeds" that ultimately grow into new autonomous civilizations (i.e., *daughter-civilizations*) that differ from their parent-civilization.

The reasons for a colony to grow into a distinctive autonomous civilization can be divided into two categories: socio-economic reasons (i.e., ownership, exploration, and control of resources, as well as modifications of language, and emergence of a unique history and culture) and reasons relevant to cosmic and planetary conditions and environments. The second category includes the unique environments and orbital parameters of colonized planets and moons, properties of their host stars, properties of interplanetary environments and interstellar environments. For example, a unique combination of the surface gravity and physical environments of a colonized planet (or a moon) would necessitate artificially produced modifications of colonists' anatomy and physiology, making them better adapted to living on the colonized planet or moon. In the process, they would become different from the same species inhabiting other cosmic worlds.

Shaped by its own unique circumstances, cosmic and planetary environments, each daughter-civilization may eventually assert its distinctiveness and autonomy. In this way, the parent-civilization may create unique and autonomous daughter-civilizations inhabiting different planets, moons, or regions of space.

These considerations may apply to colonies of extraterrestrial civilizations and any future human colonies in the Solar System and beyond. These considerations may not apply to extraterrestrial civilizations that are drastically different from humans in their ways of existence, communication, collaboration, and adaptation to different cosmic worlds.

A civilization of Cosmic Hitchhikers using a free-floating planet as interstellar transportation may establish its colonies in several planetary system when its free-floating planet passes by those planetary systems. Therefore, it may act as a 'parent-civilization' spreading the seeds of 'daughter-civilizations' in the form of its colonies in planetary systems. Its colonies may grow into autonomous civilizations, while populations of the parent-civilization remaining on the free-floating planet travel away from them.

Even if extraterrestrial civilizations engaging in Cosmic Hitchhiking were a very rare event, their space colonization could produce a considerable accumulated effect over billions of years. For example, if over the course of 10 million years, species of 1 extraterrestrial civilization would go Cosmic Hitchhiking in our Galaxy, that would equal 300 extraterrestrial civilizations sending Cosmic Hitchhikers (biological species, post-biological species, or machines) over the course of 3 billion years. The Cosmic Hitchhikers could establish multiple colonies in other planetary systems, and their colonies could grow into new civilizations. As a result, over the course of 3 billion years, Cosmic Hitchhiking of 300 extraterrestrial civilizations could lead, for example, to the rise of 900 advanced extraterrestrial civilizations (depending on how many colonies would survive).

Considering the challenges of colonization of cosmic worlds, I hypothesize that such civilizations might avoid broadcasting their existence and limit their messaging to that among parent-civilizations, their colonies, and their daughter-civilizations.

## 7. Conclusion and Recommendations

Advanced extraterrestrial civilizations may use free-floating planets as interstellar transportation for their Cosmic Hitchhikers for the purpose of space exploration and interstellar colonization. Cosmic Hitchhikers can be large groups or populations of biological species, post-biological species, and technologies. Whereas this type of interstellar travel may not save travel time, it allows space travelers to avoid technical challenges of using spacecraft for interstellar travel of large populations of species. Some Cosmic Hitchhikers might make their travel time shorter when applying astronomical engineering methods or using hypervelocity free-floating planets ejected from the central regions of the Galaxy. Extraterrestrial civilizations can also send Cosmic Hitchhikers in the form of smart technologies to ride free-floating planets and to survey interstellar space and planetary systems that the free-floating planets pass by.

Just as human scientists and engineers are looking for ways to change the motion of asteroids in the Solar System, more advanced civilizations may be able to use astronomical engineering to modify the motion of free-floating planets and get them close to the home planetary systems of such civilizations, so that large groups or populations of the Cosmic Hitchhikers could travel from their home worlds to the free-floating planets and ride them. Cosmic Hitchhikers could use astronomical engineering to steer their free-floating planets to the planetary systems of their choice.

If over the course of 10 million years, species of 1 extraterrestrial civilization would go Cosmic Hitchhiking in our Galaxy, then about 300 extraterrestrial civilizations could send Cosmic Hitchhikers (biological species, post-biological species, or machines) over the course of 3 billion years. Each civilization of Cosmic Hitchhikers could establish its colonies in more than one planetary system. Over time, the colonies could grow into independent civilizations, thus changing the total number of civilizations in the Galaxy.

Even if one assumes zero probability of free-floating planets with Cosmic Hitchhikers passing by our Solar System over the last 10 thousand years (or any other number of years), we cannot with absolute certainty rule out the possibility of at least one free-floating planet with Cosmic Hitchhikers or their artifacts passing by the Solar System over the last 4 billion years.

I propose that SETI and SETA should include the search for technosignatures and artifacts of Cosmic Hitchhikers.

## Acknowledgements

I am grateful to Dr. Lisa Kaltenegger, Dr. Dan Werthimer, and Dr. Jason T Wright for their comments on the Cosmic Hitchhikers hypothesis.

## About the Author

Irina Mullins is a Professor of Physics and Astronomy at Houston Community College. She has a M.S. in Space Physics and Astronomy from Rice University, Houston, TX. Irina Mullins writes under her maiden name, Irina K. Romanovskaya. Correspondence should be addressed to irinakromanovskaya@gmail.com or irina.mullins@hccs.edu

## Author Disclosure Statement

No competing financial interests exist.